\documentclass[a4paper,11pt]{article}
\usepackage{pos}
\usepackage{caption}
\usepackage{subcaption}
\usepackage{amsmath}
\usepackage{dsfont}
\usepackage{verbatim}
\usepackage{orcidlink}
\usepackage{graphicx}

\newcommand{\orcidauthorMASON}{0000-0002-1857-1085}
\newcommand{\orcidauthorBENNETT}{0000-0002-1678-6701}
\newcommand{\orcidauthorLUCINI}{0000-0001-8974-8266}
\newcommand{\orcidauthorPIAI}{0000-0002-2251-0111} 
\newcommand{\orcidauthorRINALDI}{0000-0003-4134-809X} 
\newcommand{\orcidauthorVADACCHINO}{0000-0002-5783-5602}
\newcommand{\orcidauthorZIERLER}{0000-0002-8670-4054}

\title{Updates on the density of states method in finite temperature symplectic gauge theories}
\ShortTitle{Density of states in $\SpN$ gauge theories}

\author*[a,f]{David Mason\,\orcidlink{\orcidauthorMASON}}
\author[c]{Ed Bennett\,\orcidlink{\orcidauthorBENNETT}}
\author[b,c]{Biagio Lucini\,\orcidlink{\orcidauthorLUCINI}}
\author[a]{Maurizio Piai\,\orcidlink{\orcidauthorPIAI}}
\author[d]{Enrico Rinaldi\,\orcidlink{\orcidauthorRINALDI}}
\author[e]{Davide Vadacchino\,\orcidlink{\orcidauthorVADACCHINO}}
\author[a]{Fabian Zierler\,\orcidlink{\orcidauthorZIERLER}}

\affiliation[a]{Department of Physics, Faculty of Science and Engineering, Swansea University (Park Campus),
Singleton Park, SA2 8PP Swansea, Wales, United Kingdom}
\affiliation[b]{Department of Mathematics, Faculty of Science and Engineering, Swansea University (Bay Campus),
Fabian Way, SA1 8EN Swansea, Wales, United Kingdom}
\affiliation[c]{Swansea Academy of Advanced Computing, Swansea University (Bay Campus), Fabian Way, SA1 8EN Swansea, Wales, United Kingdom}
\affiliation[d]{Interdisciplinary Theoretical \& Mathematical Science Program, RIKEN (iTHEMS), 2-1 Hirosawa, Wako, Saitama, 351-0198, Japan}
\affiliation[e]{Centre for Mathematical Science, University of Plymouth, Plymouth, PL4 8AA, United Kingdom}
\affiliation[f]{Jülich Supercomputing Centre, Forschungszentrum Jülich, D-52425 Jülich, Germany}

\emailAdd{2036508@Swansea.ac.uk}
\emailAdd{d.mason@fz-juelich.de}

\abstract{First-order phase transitions in the early universe have rich phenomenological implications, such as the production of a
potentially detectable signal of stochastic relic background gravitational waves. The hypothesis that new, strongly coupled dynamics, hiding in a new dark sector, could be detected in this way, via the telltale signs of its confinement/deconfinement phase transition, provides a fascinating opportunity for interdisciplinary synergy between lattice field theory and astro-particle physics. But its viability relies on completing the challenging task of providing accurate theoretical predictions for the parameters characterising the strongly coupled theory.  Density of states methods, and in particular the linear logarithmic relaxation (LLR) method, can be used to address the intrinsic numerical difficulties that arise due the meta-stable dynamics in the vicinity of the critical point. For example, it allows one to obtain accurate determinations of thermodynamic observables that are otherwise inaccessible, such as the free energy. In this contribution, we present an update on results of the analysis of the finite temperature deconfinement phase transition in a pure gauge theory with a symplectic gauge group, $Sp(4)$, by using the LLR method. We present a first analysis of the properties of the transition in the thermodynamic limit, and provide a road map for future work, including a brief preliminary discussion that will inform future publications.}

\FullConference{The 41st International Symposium on Lattice Field Theory (LATTICE2024)\\
 28 July - 3 August 2024\\
Liverpool, UK\\}
\newcommand{\beq}{\begin{equation}}
\newcommand{\eeq}{\end{equation}}

\begin{document}

\newcommand{\sun}{$SU(N_c)$~}
\newcommand{\suTN}{$SU(2N)$~}
\newcommand{\suthree}{$SU(3)$~}

\makeatletter
\newsavebox{\@brx}
\newcommand{\llangle}[1][]{\savebox{\@brx}{\(\m@th{#1\langle}\)}%
  \mathopen{\copy\@brx\mkern2mu\kern-0.9\wd\@brx\usebox{\@brx}}}
\newcommand{\rrangle}[1][]{\savebox{\@brx}{\(\m@th{#1\rangle}\)}%
  \mathclose{\copy\@brx\mkern2mu\kern-0.9\wd\@brx\usebox{\@brx}}}
\newcommand{\SpN}{S\mspace{-2mu}p(2N)}%
\newcommand{\SpF}{S\mspace{-2mu}p(4)}%
\makeatother

\maketitle
\section{Introduction}
\label{intro}

First-order phase transitions are characterised by the co-existence of phases for choices of the control parameters in proximity of the transition. This hallmark leads to interesting consequences, both phenomenologically and from a technical viewpoint. Phenomenologically, systems containing first-order phase transitions in the early universe lead to the nucleation of bubbles in the false vacuum, and the possible generation of a stochastic gravitational wave background~\cite{Witten:1984rs,Kamionkowski:1993fg,Allen:1996vm,Schwaller:2015tja,
 Croon:2018erz,Christensen:2018iqi}; see Ref.~\cite{Pasechnik:2023hwv} and references therein for further details. They can also give rise to the out-of-equilibrium condition identified by Sakharov for baryogenesis \cite{Sakharov:1967dj}. In the early universe, the Standard Model of particle physics predicts that no such phase transitions arise, as the electroweak symmetry breaking and the chiral/confinement transition are both established to be smooth cross-overs, see Refs.~\cite{DOnofrio:2015gop,Aoki:2006we}, for a recent discussion of the status of the determination in QCD, see the review~\cite{Aarts:2023vsf}.
The first detection of gravitational waves \cite{LIGOScientific:2016aoc}, and early hints pointing towards the possible detection of long wave length gravitational waves\cite{NANOGrav:2023hvm}, have driven a revival of interest in new dark sectors arising in physics beyond the standard model (BSM), and revitalised the global effort to predict and characterise  first-order phase transitions in new theories, particularly in strongly coupled ones---see for example Refs.~\cite{Huang:2020crf,Halverson:2020xpg,Kang:2021epo,Reichert:2021cvs,Reichert:2022naa,Pasechnik:2023hwv} and references therein.

Many BSM proposals contain a new strongly interacting sector, with new gauge interactions and matter fields. These new gauge sectors are motivated as possible ways to resolve some of the open questions left unanswered by the Standard Model, such as the existence of dark matter, or the naturalness problem in the electroweak theory. Interest in $Sp(2N)$ gauge theories as the basis for models to address these questions  has led to a research programme, Theoretical Explorations on the Lattice with Orthogonal and Symplectic groups (TELOS), that aims to study these theories using lattice methods, see Refs.~\cite{Bennett:2017kga,Bennett:2019jzz,Bennett:2019cxd,Bennett:2020hqd,Bennett:2020qtj,Bennett:2022yfa,Bennett:2022gdz,Bennett:2022ftz,Bennett:2023wjw,Bennett:2023gbe,Bennett:2023mhh,Bennett:2023qwx,Bennett:2024cqv,Bennett:2024wda, Lee:2018ztv,Lucini:2021xke,Bennett:2021mbw,Hsiao:2022gju,Hsiao:2022kxf}.  Gauge theories with $Sp(2N)$ group have also been studied as models of dark matter in Refs.~\cite{Maas:2021gbf,Zierler:2021cfa,
Kulkarni:2022bvh,Zierler:2022qfq,Zierler:2022uez,Bennett:2023rsl,Dengler:2023szi,Dengler:2024maq}.
This contribution reports on recent advances towards the characterisation of the finite temperature deconfinement phase transition in $Sp(4)$ pure gauge theory, using lattice field theory numerical methods.

These new gauge sectors, may be expected to undergo a deconfinement transition at the high temperatures one expects in the early universe. The properties of the transition depend on both the gauge symmetry and the matter content of the theory of interest, and their measurement is mired by a number of non-trivial challenges---see the comprehensive discussion in the introduction of Ref.~\cite{Borsanyi:2022xml},  for theories with $SU(3)$ gauge group. In Yang-Mills theories, the pure gauge dynamics of the theory, in the absence of fermion matter fields,  is believed to lead to  a 
first-order transition in all $SU(N_c)$ and $Sp(2N)$---with the notable exception of  $Sp(2) = SU(2)$---as well as in the exceptional gauge theory $G(2)$---see Refs.~\cite{Lucini:2005vg, Panero:2009tv,Holland:2003kg, Holland:2003jy, Bruno:2024dha, Cossu:2007dk} and references therein.

When studying strongly coupled field theories on the lattice, standard importance sampling approaches face a major technical limitation, when used to explore the dynamics in proximity of the phase transition. This is an intrinsic drawback arising from metastable dynamics connected with the phase co-existence around the critical point. The difficulty is that in order to obtain accurate results, one must ensure the whole phase space is accurately sampled, but the metastability introduces global energy barriers that may be challenging to overcome in local update algorithms. This may result in the configuration updates becoming stuck in one (metastable) vacuum, with tunnelling between vacua becoming exponentially suppressed, and hence may either lead to uncontrolled systematic errors, or to the demand for unrealistically large computational resources being needed to restore ergodicity. This problem becomes intractable in the infinite volume limit in which the tunnelling time between vacua diverges. Our methodology of choice, to overcome this limitation, is the adoption of the Linear Logarithmic Relaxation method (LLR), introduced in Ref.~\cite{Langfeld:2012ah}---see also Refs.~\cite{Langfeld:2013xbf,Langfeld:2015fua,Cossu:2021bgn}. We will discuss the main ideas that make this approach promising, later in this document.

The results we present here are part of an ongoing programme of study of first-order deconfinement phase transition in non-Abelian Yang-Mills theories using the LLR method. Our previous results on $SU(3)$ pure gauge theory can be found in Ref.~\cite{Lucini:2023irm}---see also Refs.~\cite{Mason:2022trc,Mason:2022aka}.
Other finite-temperature studies using this general methodology exist also for $SU(4)$~\cite{Springer:2021liy,
   Springer:2022qos} and general $SU(N)$~\cite{Springer:2023wok,Springer:2023hcc}. 
 An extended, detailed exposition of both intermediate and final results presented here can be found in Ref.~\cite{Bennett:2024bhy},
 but in this contribution we add a brief discussion on the direction of future work.

\section{First order phase transitions, lattice field theory and the LLR method}
\label{sec:Setup}

In order to study numerically the strongly coupled dynamics of the $Sp(4)$ Yang-Mills theory we use lattice field theory. The position in Euclidean spacetime, $x$, is discretised onto a hyper-cubic lattice with periodic boundary conditions in all directions. We denote the lattice size as  $\tilde V = a^4 N_t \times N_s^3$, where $a$ is the lattice spacing, and $N_t$ and $N_s$ are the number of sites in the temporal and spatial directions, respectively. The link variables, $U_{\mu}(x)$, are matrices  valued in the symplectic group $Sp(4)$, with $\mu$ the space-time direction. We denote as $U$ a generic configuration, i.e. a choice of matrices $U_{\mu}(x)$ for each $x$ and $\mu$. The standard Wilson action is
\beq
\label{eqn:Action}
S[U]\equiv\frac{6\tilde{V}}{a^4}(1-u_p[U]),
\eeq
where $u_p[U]$ is the average plaquette for a configuration, $U$. The partition function is given by
\beq
\label{eqn:PathIntegral}
Z_\beta \equiv  \int [DU_\mu]e^{-\beta S[U]},
\eeq
where $\beta$ is the gauge (inverse) coupling of the lattice theory. 

To study this system, we use the LLR method, in which we analyse the micro-canonical information through the density of states, defined as 
\beq
\rho(E) \equiv \int [DU] \delta (S[U] - E).
\eeq
The energy (action), $E$, of the system is discretised into small energy intervals, of size $\Delta_E / 2$, and the logarithm of the density of states, $\rho(E)$, is approximated as a piecewise linear function, in the small energy intervals $E_n - \Delta_E / 4 \leq E \leq  E_n + \Delta_E / 4$, with $n=1,\,\cdots,\,N$, by writing
\beq
\label{eqn:LLRRho}
\ln \rho (E) \approx a_n (E-E_n) + c_n, \quad c_n =  c_1 + \frac{\Delta_E}{4} a_1 + \frac{\Delta_E}{2} \sum_{k=2}^{n-1} a_k +
\frac{\Delta_E}{4} a_n \,.
\eeq 
The goal of this method is to calculate the coefficients, $a_n$, in a set of intervals relevant for the physical problem of interest. All other observables are then reconstructed from the knowledge of $a_n$. A detailed discussion of the technical aspects of the implementation of this strategy in our computations can be found in our earlier publications~\cite{Lucini:2023irm,Bennett:2024bhy}. For notational convenience, when quoting the interval size we trade the energy interval size for the interval in plaquette value,  $\Delta_{u_p} = a^4 \Delta_E / 6 \tilde V$.

Using this approximation for the density of states, $\rho(E)$, the canonical ensemble can be reconstructed. This allows for a determination of the plaquette distribution, $P_\beta(u_p)$, the partition function, $Z_\beta$, and the vacuum expectation value (VEV) of observables, $\langle {\cal O} \rangle$. The plaquette distribution and partition function are estimated
 by exploiting the formal definitions
\beq
\label{eqn:PlaqDistribution}
P_{\beta}(u_p)  = \frac{1}{Z_\beta}\rho(E)e^{-\beta E}|_{E=6\tilde V(1-u_p)/a^4}, \quad Z_\beta= \int dE \rho(E) e^{-\beta E}.
\eeq

\begin{figure}
    \centering
    \includegraphics[width=.45\textwidth]{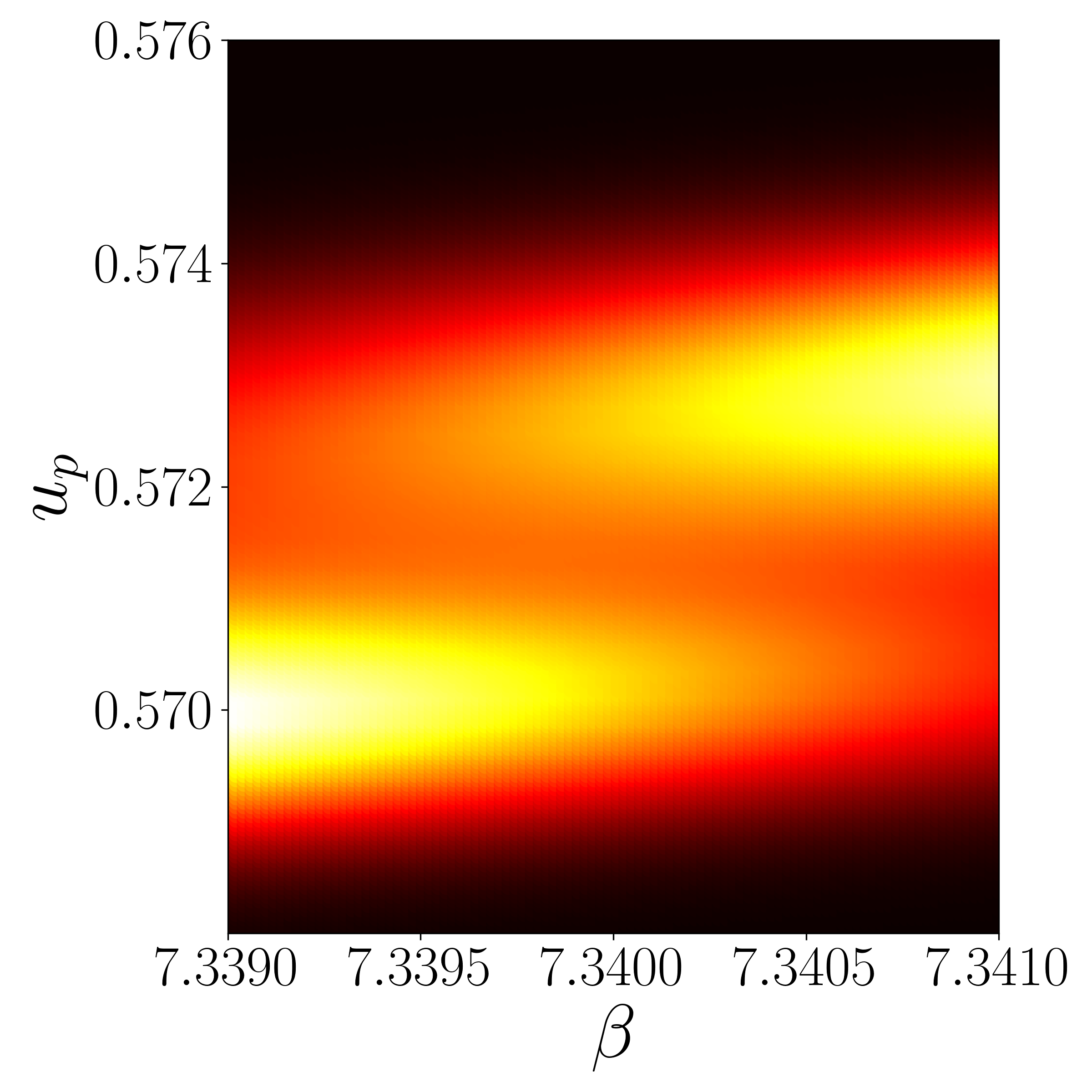}
    \includegraphics[width=.45\textwidth]{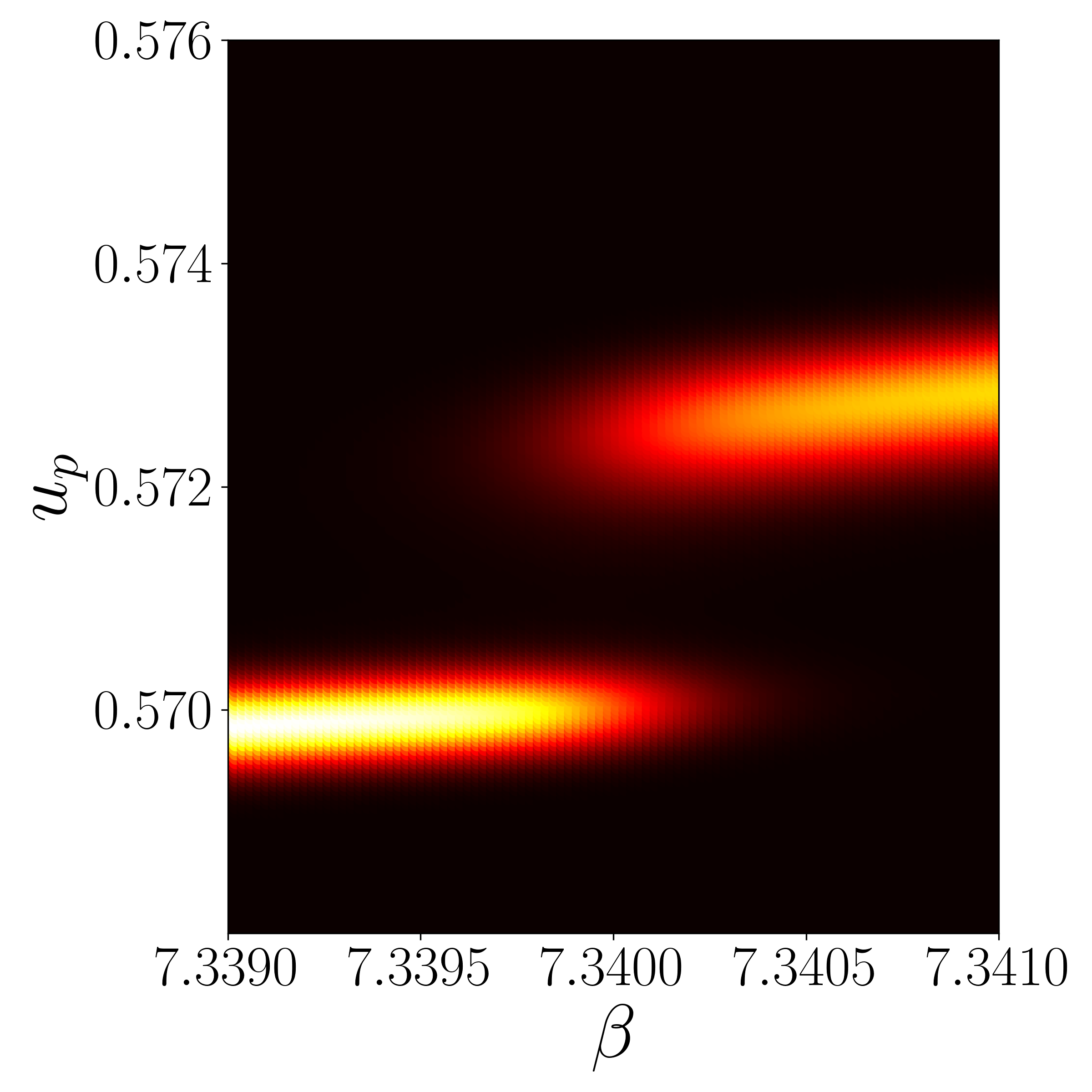}
    \caption{Colour map display of  the plaquette distribution, $P_\beta(u_p)$, for a range of couplings between $\beta = 7.339$ and $7.341$, and plaquette values between $u_p = 0.568$ and $0.576$. Brighter colours correspond to regions of phase space with higher probability. The LLR method has been applied to the  $Sp(4)$ gauge theory, on  lattices of size $4 \times 24^3$ with $\Delta_{u_p} = 0.00048$ (left panel) and $4 \times 48^3$ with $\Delta_{u_p} = 0.00013$ (right panel).  }
    \label{fig:plaqdist}
\end{figure}

A first-order phase transition can be characterised by the emergence of co-existing phases. 
Near the transition, the plaquette probability distribution displays a double-peak structure,
which is commonly approximated by a double Gaussian distribution, namely writing 
\beq
\label{eq:doublegaussianapprox}
P_\beta(u_p) = P_\beta^{(+)}(u_p) + P_\beta^{(-)}(u_p) \,, 
\eeq
where $P_\beta^{(\pm)}(u_p)$  are Gaussian distributions for states in the purely high (+) or low (-) temperature phase. The width of the individual distributions is expected to scale inversely with the volume, while the heights of the peaks should scale as the square root of the volume. At the critical point, these two distributions are of equal height. Between the two Gaussians there is 
a region of phase space with
exponentially suppressed probability. The phase transition occurs when the system tunnels from one meta-stable vacuum to another (the true vacuum). Figure~\ref{fig:plaqdist} shows the plaquette distribution for a selection of couplings chosen near the critical point, for the $Sp(4)$ pure gauge theory on lattices with size $4 \times 24^3$  and $4 \times 48^3$. 

\begin{figure}[t]
    \centering
    \includegraphics[width=.45\textwidth]{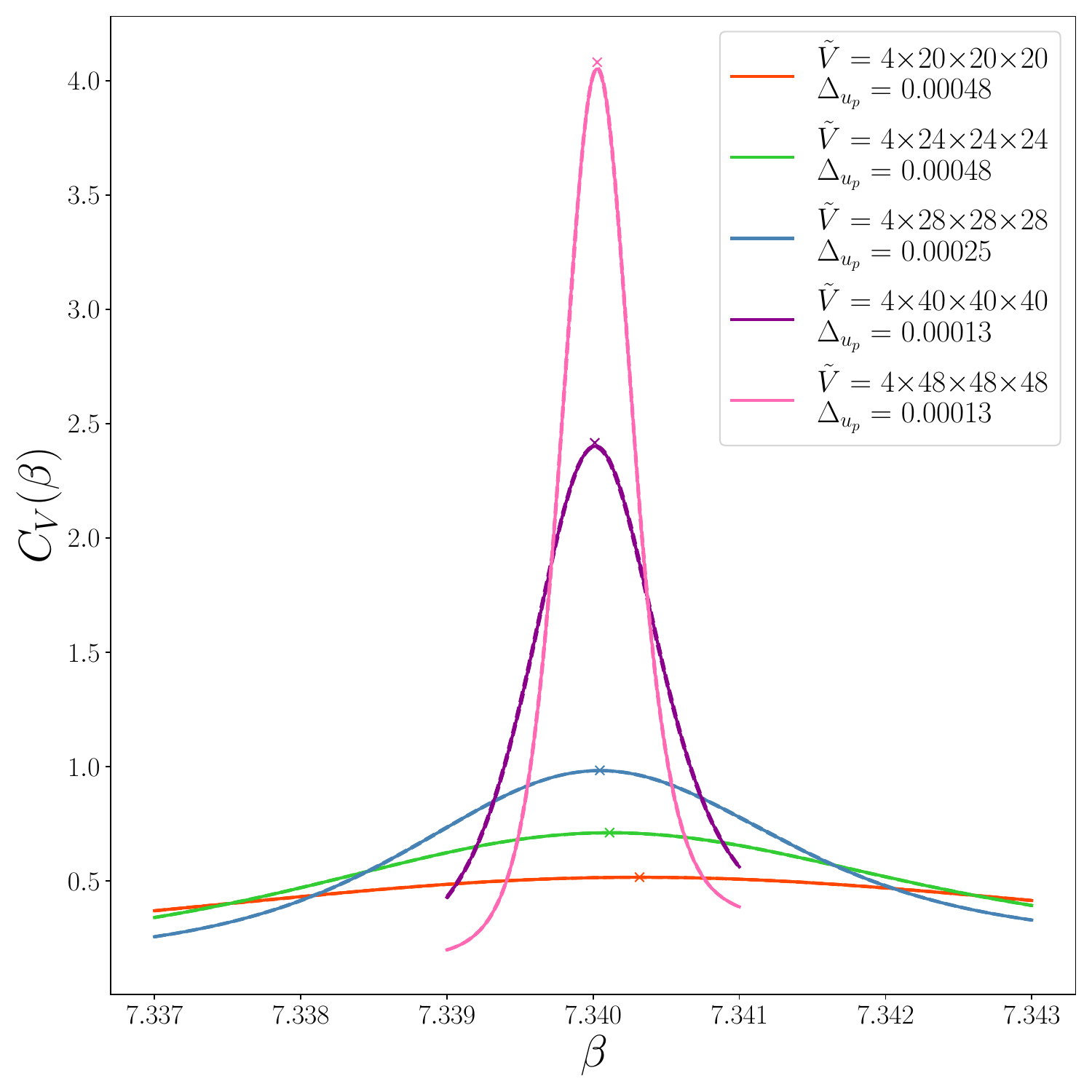}
    \includegraphics[width=.45\textwidth]{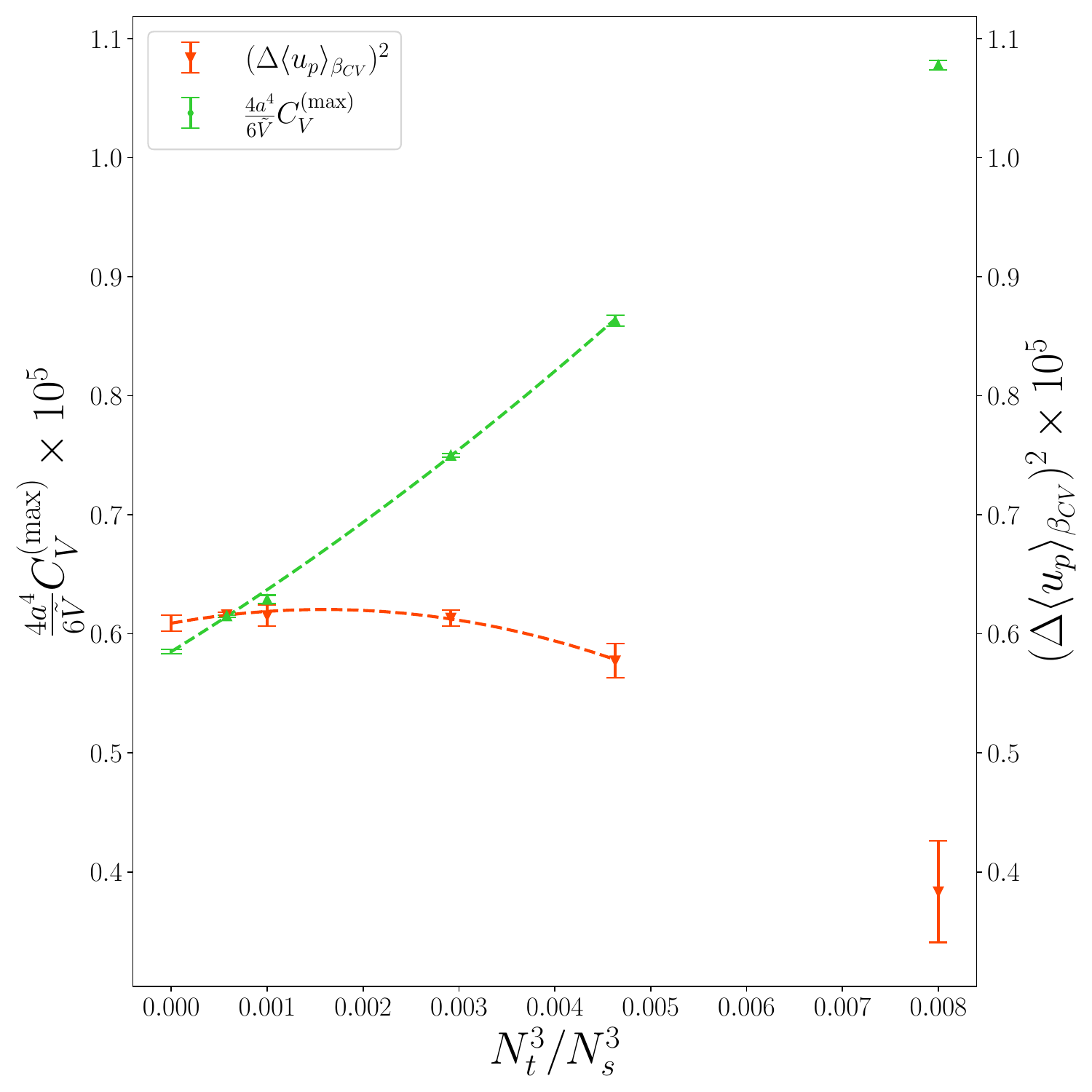}
    \caption{Left panel:  the specific heat, $C_V(\beta)$, as a function of the coupling, $\beta$, around the critical point of the $Sp(4)$ lattice gauge theory on lattice with various sizes, calculated using the LLR method. 
    Right panel:  the peak of the specific heat, $C_V^{\text{(max)}}$, normalised by a factor of $4 a^4 / 6 \tilde V$ (in green), and the square of the plaquette jump, $(\Delta \langle u_p \rangle_{\beta_{CV}})^2$  (in red), as a function of
     the cube of the inverse of the aspect ratio, $N_t^3 / N_s ^3$, of several choices of lattice, with fixed $N_t=4$. A line has been fitted to both of these results (dashed) and an extrapolation towards the thermodynamic limit is taken.}
    \label{fig:SH}
\end{figure}

The leading-order, finite-size scaling of observables at the critical point can be modelled with the guidance offered by the double Gaussian approximation. In the work presented in this contribution, we report only on the specific heat, defined as 
\begin{eqnarray}
C_V(\beta) &=& \frac{6 \tilde{V}}{a^4} \left(\langle u_p \rangle_\beta^2 - \langle u_p^2 \rangle_\beta \right)\,,
\end{eqnarray}
This quantity measures the fluctuation of the plaquette. At the critical point of a first-order phase transition, and in the thermodynamic limit, we expect the appearance of a discrete jump in the average plaquette, as the system tunnels from one meta-stable vacuum to another. This leads to a divergent specific heat, that is smoothened out by finite-volume effects away from the thermodynamic limit. The peak of the specific heat, $C_V^{\text{(max)}}$, is known to scale with the volume, $C_V^{\text{(max)}} \propto V = a^3 N_s^3$, with the error on $C_V^{\text{(max)}} / V $ being inversely proportional to the volume~\cite{Challa:1986sk}.     

To verify the leading-order scaling of the peak of the specific heat, and its relation to the plaquette jump in the thermodynamic limit, we calculate both quantities for a selection of lattice sizes, with $N_t\times N_s^3=4 \times 20^3,\, 4 \times 24^3,\,
4 \times 28^3,\, 4 \times 40^3,$ and $4 \times 48^3$. The left panel of Fig.~\ref{fig:SH} shows the specific heat calculated
using the LLR method in the proximity of the transition.
For each choice of lattice volume, $V$, these results are calculated  with one interval size, $\Delta u_p$. The figure shows how, as the volume increases, the width of the peak decreases, and its height increases.

The discontinuity  in the value of the plaquette at the critical point is related to an important thermodynamic quantity, the latent heat. At finite volume, the smoothing of the VEV due to finite-size effects makes it difficult to directly determine the size of the jump. In the thermodynamic limit, though, it is equivalent to the difference between the peaks of the plaquette distribution at the critical point, $\Delta \langle u_p \rangle_{\beta_{CV}}$. In the infinite volume limit, in which the two meta-stable peaks are expected to a become Dirac delta distributions, the peak of the specific heat can be related to the plaquette discontinuity through the relation
\begin{eqnarray}
\lim_{\tilde V/a^4 \to \infty}\frac{a^4}{\tilde V}C_V^{\text{(max)}} \to \frac{3}{2} (\Delta \langle u_p \rangle_{\beta_{CV}})^2\,.\end{eqnarray}

\begin{figure}[t]
    \centering
    \includegraphics[width=.45\textwidth]{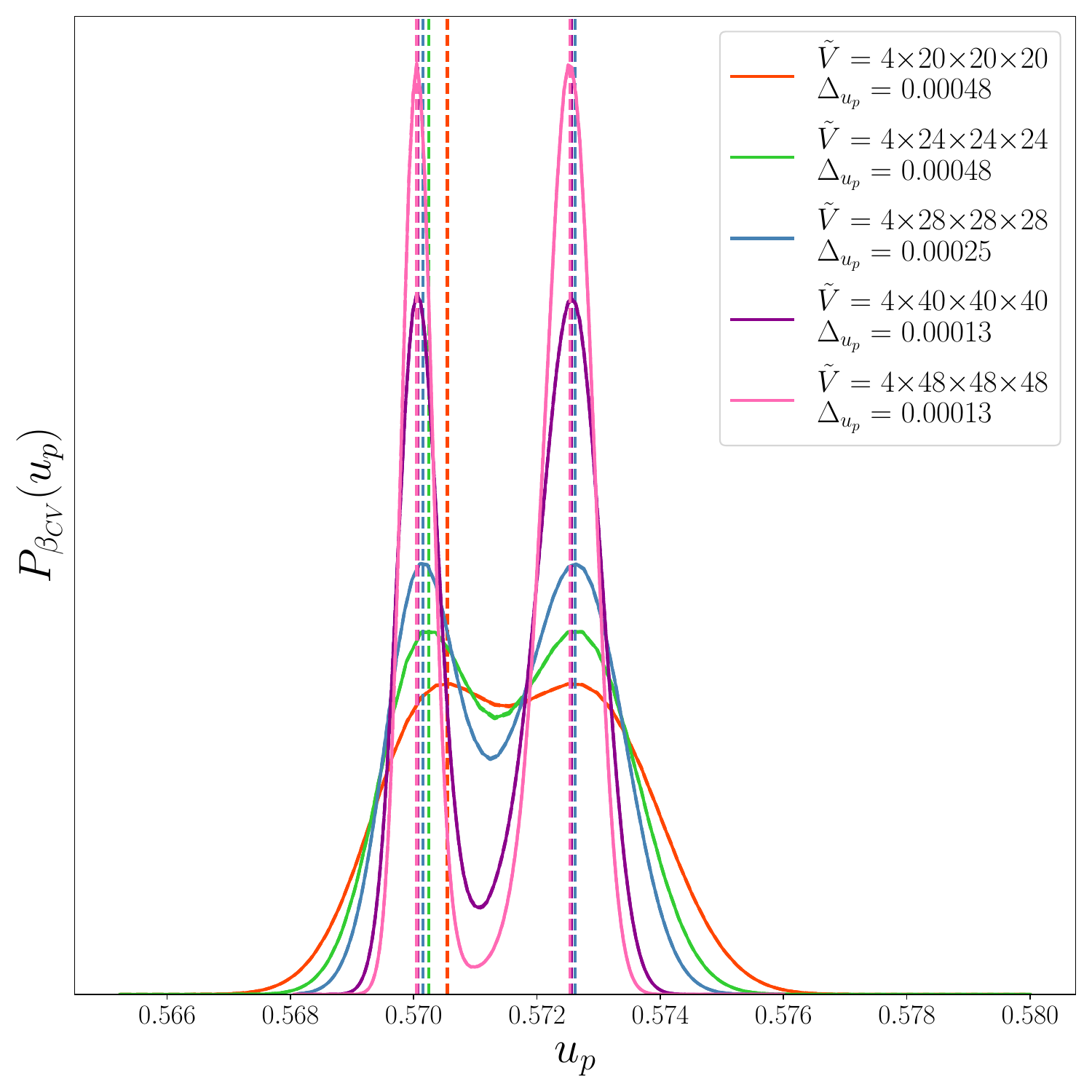}
    \includegraphics[width=.45\textwidth]{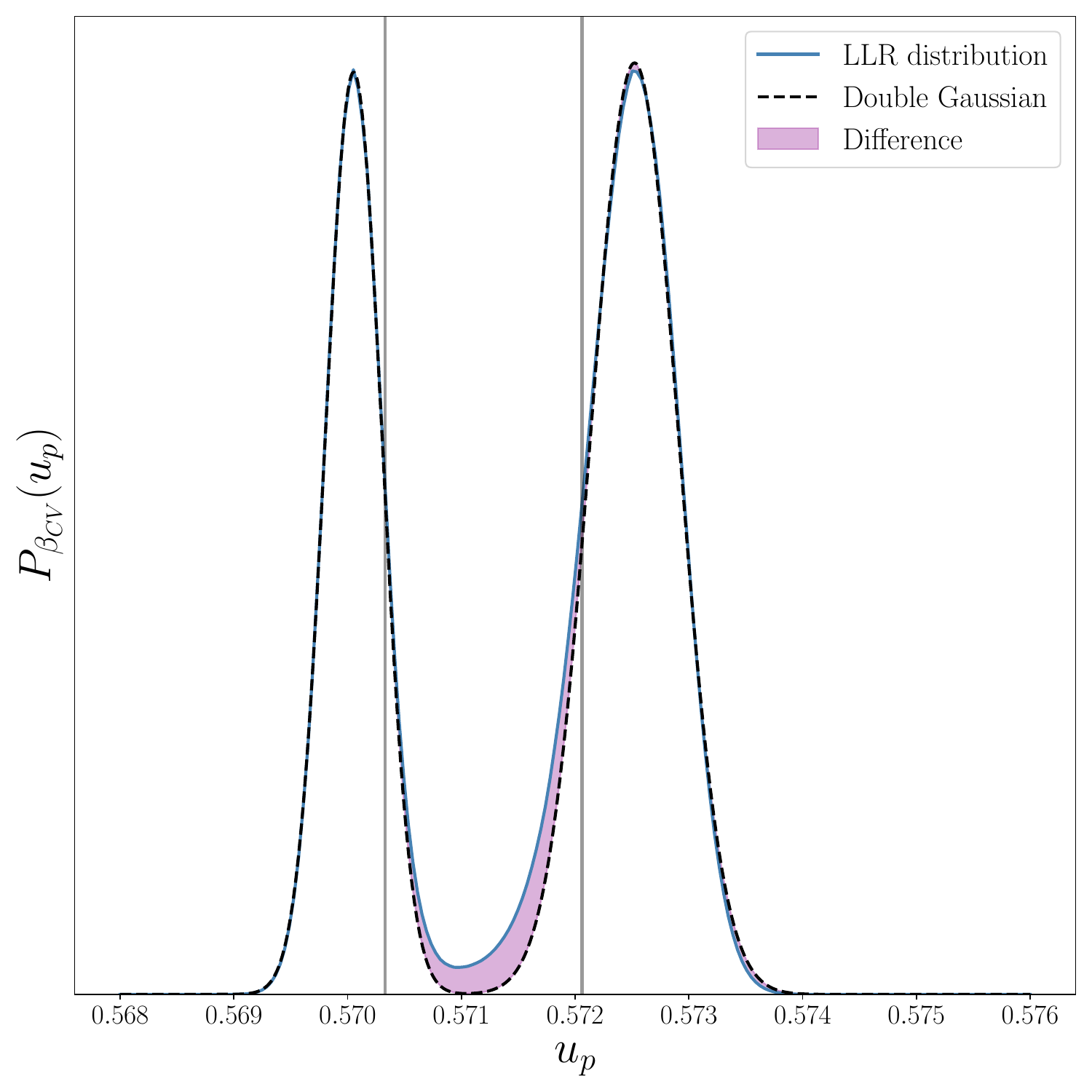}
    \caption{Left panel: the plaquette distribution at the critical point, reconstructed using the LLR method for the $Sp(4)$ lattice gauge theory on a selection of volumes, computed with finite interval size. The dashed lines show the locations of the peaks of the distribution. Right panel: the plaquette distribution of the largest available lattice, $N_t\times N_s^3=4\times 48^3$, evaluated at the critical point (blue continuous line), compared with  a double Gaussian distribution fit of the same measurements (black dashed line), obtained by including in the fit region the measurements falling
 outside the region delimited by the two  vertical black lines. The difference between the two is shown by the magenta area.}
    \label{fig:Pb_c}
\end{figure}

In the right panel of Fig.~\ref{fig:SH}, the peak of the specific heat, normalised by a factor $4 a ^4 / 6\tilde V$, is plotted as a function of the cube of the inverse of the aspect ratio, $N_t ^ 3 / N_s ^3$. We also display, on the same plot, the discontinuity of the plaquette. The limit of vanishing interval size has been taken, as discussed in the appendices of Ref.~\cite{Bennett:2024bhy}. For both sets of results, we provide a fit with a simple quadratic polynomial in $(N_t^3 / N_s^3)$,  and superimpose the resulting best fit curve to the measurements. The reduced $\chi$-square for the fit of the specific heat is $\chi^2/N_{\rm d.o.f.} = 5.5$. The  extrapolation to the thermodynamic limit gives by $\lim_{(N_t^3 / N_s^3) \to 0}(4 a^4 / 6 \tilde V) C_V^{\text{(max)}} = 5.85(2)\times 10 ^{-6}$. For the plaquette jump $\chi^2/N_{\rm d.o.f.} = 0.18$ and $\lim_{(N_t^3 / N_s^3) \to 0} (\Delta \langle u_p \rangle_{\beta_{CV}})^2 = 6.09(7)\times 10 ^{-6}$.

The figure shows that at finite volume the measurements of $\frac{a^4}{\tilde V}C_V^{\text{(max)}} $ and $ \frac{3}{2} (\Delta \langle u_p \rangle_{\beta_{CV}})^2$ are inconsistent with one another, as expected. They approach one another in the thermodynamic limit, but with two very different functional dependences on the volume $V\propto N_s^3$.
However, ultimately the numerical results we have, for $N_t=4$, display a statistically significant
discrepancy  in the thermodynamic limit.
On the one hand,  the values of the reduced $\chi$-square we reported  for the fit to the specific heat indicates that other effects may need to be taken into account. In this respect, one should highlight that the measurements we report have remarkably small statistical errors, requiring a new level of control over the systematics. For example, the large value of the $\chi^2$ might be an indication that we are underestimating the statistical errors, and the degree of correlation in the LLR measurements might need a further reassessment.

 On the other hand, there is some evidence in our measurements  that one of such systematic effects might have a very important 
 and interesting physical meaning. Although the double Gaussian approximation is a good leading-order approximation, it neglects a contribution that becomes particularly important when one tries to model the (infinite volume) thermodynamics at the transition. The presence of mixed phase configurations, for example a low temperature bubble within a high temperature background, can provide a contribution to the plaquette distributions that is approximately flat, in the region between the two peaks. This plateau becomes more clear at larger volumes, as the contributions from the pure-phase configurations are more suppressed between the two peaks. If this effect is substantial indicating the presence of a non-trivial contribution of the thermodynamics of this additional mixed phase, then the method of extrapolation to the thermodynamic limit may need to be revisited.
 
 To test this interesting hypothesis, we show the plaquette distribution at the critical point, at which the two pure phase contributions to the plaquette distribution have peaks of equal height,  in the left panel of Fig.~\ref{fig:Pb_c}, for a few choices of lattice volumes. The dashed vertical lines indicate the location of the peaks of the distribution. The spacing between the two provides a measurement of the plaquette discontinuity. It can be seen that as the volume grows the peaks become narrower, and their height grows. We can also see that the region between the peaks never vanishes. In the right panel of this plot we fit a double Gaussian to the plaquette distribution at the critical point for lattices with $N_t\times N_s^3=4 \times 48^3$. We see that the fitted double Gaussian vanishes in the unstable region, while the observed plaquette distribution plateaus. This difference demonstrates the importance of the mixed phase states, which requires a refinement of the infinite volume extrapolations.


\section{Summary and Outlook}

Using the LLR method, we have computed the plaquette distribution, the peak of the specific heat, and the plaquette discontinuity, evaluated at the critical point for the deconfinement phase transition, in the $Sp(4)$ pure lattice gauge theory, for a range of volumes, $N_t\times N_s^3=4 \times 20^3,\, 4 \times 24^3,\, 4 \times 28^3,\, 4 \times 40^3$, and $4 \times 48^3$. We have showed that at the critical point, for the largest volumes available, the plaquette distribution starts to show evidence of a  systematic discrepancy with the double Gaussian approximation.  In the region between the two peaks, the discrepancy indicates the presence of a contribution to the thermodynamics coming from the mixed-phase configurations. This, combined with the evidence for a high value of the reduced $\chi$-square in the fit for the peak of specific heat, accompanied by the disagreement with the expected relation to the square of the plaquette discontinuity in the thermodynamic limit, provides intriguing evidence  that the mixed phase configurations are effecting the extrapolations and must be hence included in future analysis, for accurate results.

Since the presentation at the LATTICE2024 conference, our collaboration has continued to perform further numerical calculations within this lattice theory. We have been extending the set of lattice sizes presented here, and in Ref.~\cite{Bennett:2024bhy}, by an additional one with larger volume $N_t\times N_s^3=4 \times 80^3$, with the intention of providing more numerical data to study the deviation from the double Gaussian approximation. Unsurprisingly, our preliminary data suggests that the iterative procedure to solve for the values of $a_n$ requires more iteration steps than on smaller lattices. We are further investigating the system at larger temporal extents, $N_t=5,6$, to the purpose of performing the first  study of  the continuum limit for this theory (at finite temperature).

\acknowledgments
The work of E.~B. has been funded by the UKRI Science and Technologies Facilities Council (STFC) Research Software Engineering Fellowship EP/V052489/1, by the STFC under Consolidated Grant No. ST/X000648/1, and by the ExaTEPP project EP/X017168/1. 
The work of D.~V. is supported by STFC under Consolidated Grant No.~ST/X000680/1.
F.~Z. is supported by the STFC Consolidated Grant No.~ST/X000648/1.
The work of D.~M. is supported by a studentship awarded by the Data Intensive Centre for Doctoral Training, which is funded by the STFC grant ST/P006779/1.  
B.~L. and M.~P. received funding from the European Research Council (ERC) under the European Union’s Horizon 2020 research and innovation program under Grant Agreement No.~813942, and by STFC under Consolidated Grants No. ST/P00055X/1, ST/T000813/1, and ST/X000648/1. 
The work of B.~L. is further supported in part by the Royal Society Wolfson Research Merit Award WM170010 and by the Leverhulme Trust Research Fellowship No. RF-2020-4619. 

Numerical simulations have been performed on the Swansea SUNBIRD cluster (part of the Supercomputing Wales project) and AccelerateAI A100 GPU system.The Swansea SUNBIRD system and AccelerateAI are part funded by the European Regional Development Fund (ERDF) via Welsh Government. 
Numerical simulations have been  performed 
on the DiRAC Data Intensive service at Leicester. The DiRAC Data Intensive service equipment at Leicester was funded by BEIS capital funding via STFC capital grants ST/K000373/1 and ST/R002363/1 and STFC DiRAC Operations grant ST/R001014/1. 
Numerical simulations have used the DiRAC Extreme Scaling service at the University of Edinburgh. The DiRAC Data Intensive service at Leicester was operated by the University of Leicester IT Services, and the DiRAC Extreme Scaling service is operated by the Edinburgh Parallel Computing Centre, they form part of the STFC DiRAC HPC Facility (www.dirac.ac.uk). 
This work used the DiRAC Data Intensive service (CSD3) at the University of Cambridge, managed by the University of Cambridge University Information Services on behalf of the STFC DiRAC HPC Facility (www.dirac.ac.uk). The DiRAC component of CSD3 at Cambridge wass funded by BEIS, UKRI and STFC capital funding and STFC operations grants. DiRAC is part of the UKRI Digital Research Infrastructure.
ST/R00238X/1The DiRAC Extreme Scaling service was funded by BEIS capital funding via STFC capital grant ST/R00238X/1 and STFC DiRAC Operations grant ST/R001006/1.
DiRAC is part of the National e-Infrastructure.

This work was supported by the Supercomputer Fugaku Start-up Utilization Program of RIKEN.
This work used computational resources of the supercomputer Fugaku provided by RIKEN through the HPCI System Research Project (Project ID: hp230397).

{\bf Open Access Statement - } For the purpose of open access, the authors have applied a Creative Commons 
Attribution (CC BY) licence  to any Author Accepted Manuscript version arising.

{\bf Research Data Access Statement}--- The results presented in this manuscript are expanded in Ref.~\cite{Bennett:2024bhy}, the data and analysis code for this a larger publication can be downloaded from  Ref.~\cite{DataRelease1}. The simulation code can be found from Ref.~\cite{DataRelease2}.  
\bibliographystyle{jhep}
\bibliography{references}
\end{document}